\documentclass[twocolumn,showpacs,preprintnumbers,amsmath,amssymb,prb]{revtex4}

\usepackage{graphicx}
\usepackage{dcolumn}
\usepackage{bm}
\usepackage{color}

\begin{document}

\preprint{}

\title{Multiple lattice instabilities resolved by magnetic-field and disorder sensitivities in cubic paramagnetic phase of the orbital-degenerate frustrated spinel MgV$_2$O$_4$}

\author{Tadataka Watanabe$^1$}
\author{Takashi Ishikawa$^1$}
\author{Shigeo Hara$^2$}
\author{A. T. M. Nazmul Islam$^3$}
\author{Elisa M. Wheeler$^3$}
\author{Bella Lake$^3$}
\affiliation{$^1$Department of Physics, College of Science and Technology (CST), Nihon University, Chiyoda, Tokyo 101-8308, Japan}
\affiliation{$^2$Department of Physics, Chuo University, Bunkyo, Tokyo 101-8324, Japan}
\affiliation{$^3$Helmholtz-Zentrum Berlin f\"ur Materialien und Energie, Hahn-Meitner-Platz 1, D-14109 Berlin, Germany}
\date{\today}

\begin{abstract}
Ultrasound velocity measurements of the orbital-degenerate frustrated spinel MgV$_2$O$_4$ are performed in the high-purity single crystal which exhibits successive structural and antiferromagnetic phase transitions, and in the disorder-introduced single crystal which exhibits spin-glass-like behavior. The measurements reveal that two-types of unusual temperature dependence of the elastic moduli coexist in the cubic paramagnetic phase, which are resolved by magnetic-field and disorder sensitivities: huge Curie-type softening with decreasing temperature, and concave temperature dependence with a characteristic minimum. These elastic anomalies suggest the coupling of lattice to coexisting orbital fluctuations and orbital-spin-coupled excitations.
\end{abstract}


\maketitle

Vanadate spinels $A$V$_2$O$_4$ with divalent $A^{2+}$ ions such as Zn$^{2+}$, Mg$^{2+}$, Cd$^{2+}$, and Mn$^{2+}$ have attracted considerable interest as orbital-degenerate frustrated magnets, where the trivalent magnetic V$^{3+}$ ions are characterized by double occupancy of the triply-degenerate $t_{2g}$ orbitals, and form a sublattice of corner-sharing tetrahedra \cite{Radaelli}. These spinels undergo successive structural and magnetic phase transitions: a cubic-to-tetragonal lattice compression at a temperature $T_s$, and an antiferromagnetic (AF) ordering at a lower temperature $T_N$, $T_s>T_N$, for nonmagnetic $A$ = Zn \cite{Lee2}, Mg \cite{Mamiya}, and Cd \cite{Giovannetti} (a ferrimagnetic ordering at a higher temperature $T_c$, $T_s<T_c$, for magnetic $A$ = Mn \cite{Adachi}).

The possibility of novel orbital and magnetic orderings in $A$V$_2$O$_4$ has been extensively discussed both theoretically \cite{Tsunetsugu,Tchernyshyov,Matteo,Marita,Kaur,Sarkar,Pardo} and experimentally \cite{Lee2,Wheeler,Suzuki,Garlea} which are considered to be driven due to the competition of Jahn-Teller coupling, Kugel-Khomskii exchange interaction \cite{Kugel}, and relativistic spin-orbit coupling. From the discussion on the orbital and magnetic orders in $A$V$_2$O$_4$, it is considered that the absolute and relative magnitude of these three interactions in $A$V$_2$O$_4$ differs from compound to compound. For instance, it has been pointed out that the effect of the spin-orbit coupling on the orbital-ordered structure is strong in ZnV$_2$O$_4$ while negligibly weak in MnV$_2$O$_4$ \cite{Marita,Sarkar}.

In this paper, we study the interplay of orbital, spin, and lattice degrees of freedom in the magnesium vanadate spinel MgV$_2$O$_4$ ($T_s$ = 65 K and $T_N$ = 42 K) by means of ultrasound velocity measurements. The sound velocity or the elastic modulus is a useful probe which can extract symmetry-resolved thermodynamic information in the frustrated magnets \cite{Watanabe1,Watanabe2}. In MnV$_2$O$_4$, recent ultrasound velocity measurements observed a huge elastic softening on cooling in the cubic paramagnetic (PM) phase, which is considered to be driven by the coupling of lattice to orbital-spin-coupled fluctuations \cite{Nii}.

For MgV$_2$O$_4$, it was suggested from the inelastic neutron scattering experiments that the effect of the spin-orbit coupling on the orbital-ordered structure is stronger than that in MnV$_2$O$_4$, but weaker than that in ZnV$_2$O$_4$ \cite{Wheeler}. Thus the orbital and spin states of MgV$_2$O$_4$ are expected to vary from those of MnV$_2$O$_4$ and ZnV$_2$O$_4$ \cite{Kaur}. The present study particularly focuses on the orbital and spin states in the orbital- and spin-disordered phase (the cubic PM phase) of MgV$_2$O$_4$.

The ultrasound velocity measurements were performed on two different types of MgV$_2$O$_4$ single crystals grown by the floating-zone method: one is a high-purity single crystal which exhibits a cubic-to-tetragonal structural transition at $T_s$ = 65 K and an AF transition at $T_N$ = 42 K, named here as "ordered MgV$_2$O$_4$", and the other is a single crystal which exhibits spin-glass-like behavior below $T_f$ = 12.5 K, named here as "disordered MgV$_2$O$_4$" \cite{Islam}. It is known that, for MgV$_2$O$_4$, a small amount of disorder suppresses the structural and magnetic phase transitions, and induces spin-glass-like behavior in low temperatures \cite{Islam}. The "ordered MgV$_2$O$_4$" is a disorder-free single crystal grown by the traveling-solvent floating-zone method, while the "disordered MgV$_2$O$_4$" is a single crystal in which $\sim$3 $\%$ of V atoms in the octahedral sites are substituted by Mg atoms \cite{Islam}. The ultrasound velocities were measured using the phase comparison technique with longitudinal and transverse sound waves at a frequency of 30 MHz. The ultrasounds were generated and detected by LiNbO$_3$ transducers glued on the parallel mirror surfaces of the crystal. We measured sound velocities in all the symmetrically independent elastic moduli in the cubic crystal: compression modulus $C_{11}$, tetragonal shear modulus $\frac{C_{11}-C_{12}}{2}\equiv C_t$, and trigonal shear modulus $C_{44}$. In the following, we denote the elastic modulus $C_{\Gamma}$ of the ordered and the disordered MgV$_2$O$_4$ by $C_{\Gamma}^o$ and $C_{\Gamma}^d$, respectively.

Figures 1(a)-1(c), respectively, depict the longitudinal sound velocity $v_L$ in $C_{11}^o$ and the transverse sound velocity $v_T$ in $C_t^o$ and $C_{44}^o$ as functions of temperature ($T$) with zero magnetic field ($H$ = 0). All the elastic modes exhibit a jump at $T_s$ and a discontinuous change at $T_N$, as marked by arrows in Figs. 1(a)-1(c). In the cubic PM phase of the ordered MgV$_2$O$_4$, $T>T_s$, all the elastic modes exhibit softening with decreasing $T$. It is noted here that the softening observed in the ordered MgV$_2$O$_4$ is divided into two types: Curie-type ($\sim-1/T$-type) softening (convex $T$ dependence) in $C_{11}^o(T)$ and $C_t^o(T)$, and nonmonotonic softening with concave $T$ dependence in $C_{44}^o(T)$, which is different from the observation of only the Curie-type softening in MnV$_2$O$_4$ \cite{Nii}.

\begin{figure}[t]
\begin{center}
\includegraphics[scale=0.33]{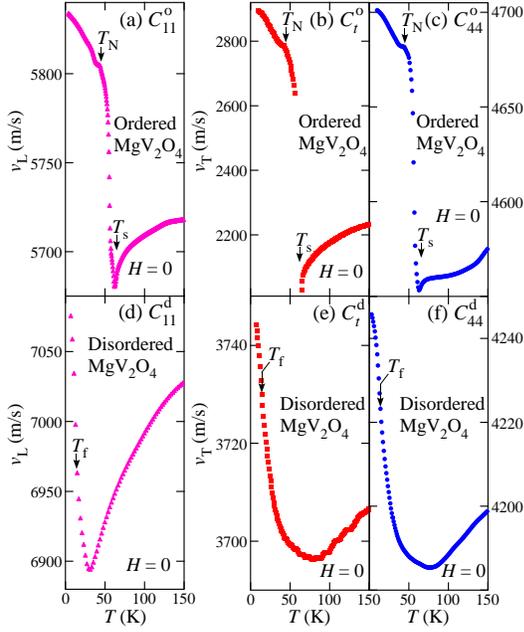}
\caption{\label{fig:Fig1} (Color online). $T$ dependence of $v_L$ and $v_T$ with $H$ = 0 in MgV$_2$O$_4$. (a)-(c) The ordered MgV$_2$O$_4$: (a) $C_{11}^o$, (b) $\frac{(C_{11}^o-C_{12}^o)}{2}\equiv C_t^o$, and (c) $C_{44}^o$. $T_s$ and $T_N$ are marked by arrows in (a)-(c). (d)-(f) The disordered MgV$_2$O$_4$: (d) $C_{11}^d$, (e) $C_t^d$, and (f) $C_{44}^d$. $T_f$ is marked by arrows in (d)-(f).}
\end{center}
\end{figure}

The $T$ dependence of $v_L$ and $v_T$ with $H$ = 0 in the disordered MgV$_2$O$_4$ behaves differently from that in the ordered MgV$_2$O$_4$, as shown in Figs. 1(d)-1(f). In the cubic PM phase of the disordered MgV$_2$O$_4$, $T>T_f$, all the elastic modes exhibit softening with decreasing $T$ but turn to hardening in low $T>T_f$. The minimum of the elastic moduli is at $\sim$30 K for $C_{11}^d(T)$, and at $\sim$ 80 K for $C_t^d(T)$ and $C_{44}^d(T)$, respectively. At $T_f$ marked by arrows in Figs. 1(d)-1(f), all the elastic modes exhibit a small increase in slope.

Figures 2(a)-2(c), respectively, depict $C_{11}^o(T)$, $C_t^o(T)$, and $C_{44}^o(T)$ with $H||[110]$ in the cubic PM phase, $T>T_s$. It is evident that the softening in $C_{11}^o(T)$ and $C_{44}^o(T)$ exhibits $H$ dependence, while the Curie-type softening in $C_t^o(T)$ is independent of $H$. In particular, the softening in $C_{44}^o(T)$ is very sensitive to $H$: the nonmonotonic softening with concave curvature in the $H$ = 0 data becomes closer to the Curie-type softening in the 7 T data. Thus, taking into account that all the data shown in Fig. 2(c), $C_{44}^o(T)$ with $H$ = 0, 3 T, and 7 T, exhibit the Curie-type softening below $\sim$ 80 K, the nonmonotonic softening in $C_{44}^o(T)$ should observe a superposition of $H$-sensitive concave $T$ dependence and $H$-insensitive Curie-type softening. The weakly $H$-dependent softening in $C_{11}^o(T)$ shown in Fig. 2(a) should also observe the similar kind of superposition, although, for $C_{11}^o(T)$, the $H$-insensitive Curie-type softening is a dominant component compared to the $H$-sensitive concave $T$ dependence. For the disordered MgV$_2$O$_4$, as shown in Figs. 2(d)-2(f), $T$ dependence of all the elastic moduli with $H||[110]$ exhibits the softening with minimum on cooling in the cubic PM phase, $T>T_f$, which is insensitive to $H$.

From now on, we shall discuss the origins of the unusual elastic softening observed in the cubic PM phase of the ordered and the disordered MgV$_2$O$_4$. First we address the origin of the $H$-insensitive Curie-type softening in $C_{t}^o(T)$ and $C_{11}^o(T)$ shown in Figs. 2(b) and 2(a), respectively. Taking into account that the relative change in softening is much larger in $C_t^o(T)$ ($\frac{\Delta C_t^o}{C_t^o}\sim$15$\%$) than in $C_{11}^o(T)$ ($\frac{\Delta C_{11}^o}{C_{11}^o}\sim$1$\%$), and that $C_{11}$ is written as $C_{11}= C_B+\frac{4}{3}C_t$ with $C_B$ the bulk modulus, the Curie-type softening should be characterized as a softening in the tetragonal shear modulus $C_t^o(T)$. Thus the Curie-type softening should be a precursor to the cubic-to-tetragonal structural transition at $T_s$.

In paramagnets with the presence of some kind of magnetic fluctuations/excitations, the elastic constant $C_{\Gamma}(T)$ is generally written in the mean-field framework as \cite{Watanabe1,Luthi,Wolf}
\begin{equation}
C_{\Gamma}(T)=C_{0,\Gamma}-G_{\Gamma}^2N\frac{\chi_{\Gamma}(T)}{(1-K_{\Gamma}\chi_{\Gamma}(T))},
\label{eq:SM}
\end{equation}
with $C_{0,\Gamma}$ the background elastic constant, $N$ the density of magnetic units that generate the fluctuations/excitations, $G_{\Gamma}$ the coupling constant to a single magnetic unit, $K_{\Gamma}$ the inter-magnetic-unit interaction, and $\chi_{\Gamma}=G_{\Gamma}^{-2}(\partial^2 F/\partial \epsilon_{\Gamma}^2)$ the strain susceptibility of a single magnetic unit.

In Jahn-Teller (JT) magnets, the degenerate ground state is considered to couple strongly and selectively to the elastic modulus $C_{\Gamma}$ which has the same symmetry as the JT distortion. For such a JT-active elastic mode, assuming the single JT ion to be the magnetic unit in Eq. (1), the strain susceptibility is dominated by the Curie term $\chi_{\Gamma}(T)\sim1/T$ at low temperatures. Then Eq. (1) is rewritten as \cite{Watanabe1,Nii,Luthi}
\begin{equation}
C_{\Gamma}(T) = C_{0,\Gamma} (1-\frac{E_{JT}}{T-\theta}),
\label{eq:JT}
\end{equation}
with $C_{0,\Gamma}$ the elastic constant without JT effect, $\theta$ the intersite orbital-orbital (quadrupole-quadrupole) interaction, and $E_{JT}$ the JT coupling energy. Here $\theta$ is positive (negative) when the interaction is ferrodistortive (antiferrodistortive). In Fig. 2(b), a fit of the zero-field experimental $C_t^o(T)$ to Eq. (2) with the fit parameters listed in the right-hand side of Fig. 2(b) is depicted as a solid curve which reproduces very well the experimental data. The positive fitted value of $\theta$ = 15 K indicates the dominance of ferro-orbital intersite interactions.

\begin{figure}[t]
\begin{center}
\includegraphics[scale=0.33]{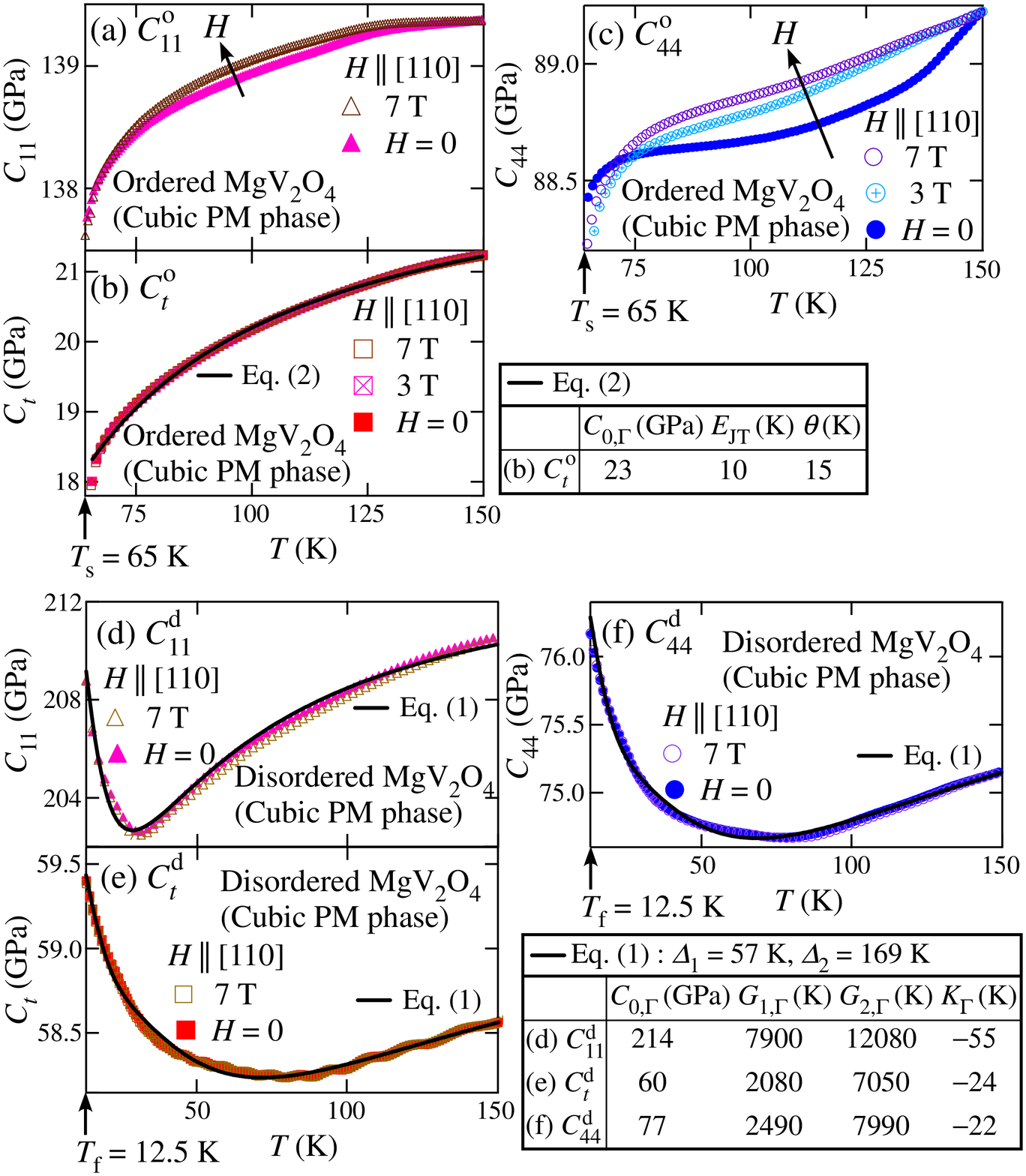}
\caption{\label{fig:Fig2} (Color online). The elastic moduli of MgV$_2$O$_4$ with $H \|$[110] as functions of $T$ in the cubic PM phase. (a)-(c) The ordered MgV$_2$O$_4$ ($T>T_s$): (a) $C_{11}^o(T)$, (b) $C_t^o(T)$, and (c) $C_{44}^o(T)$. The solid curve in (b) is a fit of the zero-field experimental $C_t^o(T)$ to Eq. (2) with the fit parameters listed in the right-hand side of (b). The solid arrows in (a) and (c) are guides to the eye indicating the variation of $C_{11}^o(T)$ and $C_{44}^o(T)$ with increasing $H$. (d)-(f) The disordered MgV$_2$O$_4$ ($T>T_f$): (d) $C_{11}^d(T)$, (e) $C_t^d(T)$, and (f) $C_{44}^d(T)$. The solid curves in (d)-(f) are fits of the zero-field experimental $C_{\Gamma}^d(T)$ to Eq. (1) with assumption of the singlet-triplet gapped excitations of the AF V$^{3+}$ clusters, respectively. The fit parameters are listed in the right-hand side of (e).}
\end{center}
\end{figure}

It should be noted here that the Curie-type softening is insensitive to $H$ in the ordered MgV$_2$O$_4$, as shown in Fig. 2(b), but sensitive to $H$ in MnV$_2$O$_4$ \cite{Nii}. For MnV$_2$O$_4$, the $H$-sensitive Curie-type softening is considered to be driven by the orbital-spin-coupled fluctuations since the Curie-type softening is enhanced in accordance with the growth of magnetization with increasing $H$. However, the magnitude of the magnetization in MgV$_2$O$_4$ is of the order of 100 times smaller than that in MnV$_2$O$_4$ \cite{Nii,Islam}. Thus the orbital fluctuations in MgV$_2$O$_4$ should be hardly affected by the spin sector, giving rise to the $H$-insensitive Curie-type softening.

Next, we discuss the origin of the softening with minimum in $C_{\Gamma}^d(T)$ shown in Figs. 2(d)-2(f). We first note that the disordered MgV$_2$O$_4$ exhibits the absence of the Curie-type softening in $C_{\Gamma}^d(T)$, namely the absence of a precursor to structural transition, which is compatible with the absence of structural transition in the disordered MgV$_2$O$_4$. Taking into account the presence of the Curie-type softening in $C_t^o(T)$ shown in Fig. 2(b), the present study reveals that not only the structural transition but also its precursor (the Curie-type softening) in MgV$_2$O$_4$ is sensitively suppressed by disorder.

As shown in Figs. 2(d)-2(f), the disordered MgV$_2$O$_4$ exhibits the $H$-insensitive softening with minimum in $C_{\Gamma}^d(T)$. According to Eq. (1), the softening with minimum in $C_{\Gamma}(T)$ is generally driven by the presence of a finite gap for the local magnetic excitations which is sensitive to the strain. Indeed, the experimental data of $C_{\Gamma}(T)$ in the spin-dimer systems such as SrCu$_2$(BO$_3$)$_2$ and the spin-frustrated systems such as MgCr$_2$O$_4$, for instance, are explained well by the quantitative analyses using Eq. (1) with assumption of the gapped excitations \cite{Wolf,Watanabe1}. The softening with minimum in $C_{\Gamma}^d(T)$ should also be attributed to the coupling of lattice to the gapped magnetic excitations.

For the spin-frustrated MgCr$_2$O$_4$, the softening with minimum observed in $C_{\Gamma}(T)$ was quantitatively analyzed using Eq. (1) with assumption of the cluster-spin excitations, where the magnetic unit in Eq. (1) is the AF hexagonal spin cluster in the Cr$^{3+}$ pyrochlore lattice \cite{Watanabe1}. In analogy with MgCr$_2$O$_4$, we now give a quantitative analysis of the experimental $C_{\Gamma}^d(T)$ in the disordered MgV$_2$O$_4$ using Eq. (1) with assumption of the singlet-triplet excitations of AF V$^{3+}$ clusters: an excitation gap $\Delta_1$ for a single V$^{3+}$ cluster and a multi-V$^{3+}$-cluster excitation gap $\Delta_2$. The contribution of the V$^{3+}$ clusters to the elastic constant should take the form of Eq. (1) with $N$ the density of V$^{3+}$ clusters, $G_{1,\Gamma}=|\partial \Delta_1/\partial \epsilon_{\Gamma}|$ the coupling constant for a single V$^{3+}$ cluster measuring the strain ($\epsilon_{\Gamma}$) dependence of the excitation gap $\Delta_1$, $K_{\Gamma}$ the inter-V$^{3+}$-cluster interaction, and $\chi_{\Gamma}(T)$ the strain susceptibility of a single V$^{3+}$ cluster. The exact shape of the V$^{3+}$ cluster remains unclear so far. However, taking into account that the inelastic neutron scattering experiments in $A$V$_2$O$_4$ ($A$ = Zn, Mg) in the cubic PM phase observed the excitations centered around a wave vector $Q$ = 1.35$\AA^{-1}$ which is smaller than $Q$ = 1.5$\AA^{-1}$ in $A$Cr$_2$O$_4$ ($A$ = Zn, Mg) \cite{Lee1,Wheeler,Lee2,Tomiyasu}, the value of $N$ = 2.52$\times$10$^{27}$ m$^{-3}$ in Eq. (1) is fixed in a first approximation which is (1.35$\AA^{-1}$/1.5$\AA^{-1}$)$^3\simeq0.73$ times of the value of $N$ = 3.45$\times$10$^{27}$ m$^{-3}$ fixed in MgCr$_2$O$_4$ \cite{Watanabe1}. Fits of the experimental data for the disordered MgV$_2$O$_4$ to Eq. (1) are depicted in Figs. 2(d)-2(f) as solid curves, respectively. With the fit parameters listed in the right-hand side of Fig. 2(e), the fits of Eq. (1) are in excellent agreement with the experimental data, reproducing the characteristic minimum in $C_{\Gamma}^d(T)$. The $K_{\Gamma}$ values are negative in all the elastic modes, indicating that the inter-V$^{3+}$-cluster interaction is antiferrodistortive. And for all the elastic modes, the coupling constant $G_{2,\Gamma}=|\partial \Delta_2/\partial \epsilon_{\Gamma}|$ is larger than $G_{1,\Gamma} =|\partial \Delta_2/\partial \epsilon_{\Gamma}|$, indicating that the higher excitations $\Delta_2$ couple to the lattice deformation more strongly than the lowest excitations $\Delta_1$.

Lastly, we discuss the $H$-sensitive nonmonotonic softening in $C_{44}^o(T)$ shown in Fig. 2(c). As already mentioned, this softening should be a superposition of the $H$-sensitive concave $T$ dependence and the $H$-insensitive Curie-type softening. The component of the Curie-type softening in $C_{44}^o(T)$ should be a precursor to the structural transition similar to that in $C_t^o(T)$. Subtracting the component of the Curie-type softening from the observed nonmonotonic softening in $C_{44}^o(T)$, another component observed as the concave $T$ dependence should be characterized as a softening with minimum which is driven by the presence of a finite gap for the magnetic excitations. The coexistence of the $H$-insensitive Curie-type softening and the $H$-sensitive softening with minimum in $C_{44}^o(T)$ strongly suggests the coexistence of orbital fluctuations and orbital-spin-coupled excitations. Furthermore, the experimental results in the ordered and the disordered MgV$_2$O$_4$ shown in Figs. 2(a)-2(f) reveal that, for MgV$_2$O$_4$, the component of the Curie-type softening is sensitively suppressed by disorder, but the component of the softening with minimum survives disorder, as illustrated in Figs. 3(a) and 3(b).

The $H$-dependent softening in $C_{11}^o(T)= C_B^o(T)+\frac{4}{3}C_t^o(T)$ shown in Fig. 2(a) should also see the superposition of the same two components as the softening in $C_{44}^o(T)$, where the $H$-insensitive Curie-type softening is the dominant component compared to the $H$-sensitive softening with minimum. Since $C_t^o(T)$ exhibits only the $H$-insensitive Curie-type softening as shown in Fig. 2(b), the component of the $H$-sensitive softening with minimum should arise in the bulk modulus $C_B^o(T)$.

\begin{figure}[t]
\begin{center}
\includegraphics[scale=0.33]{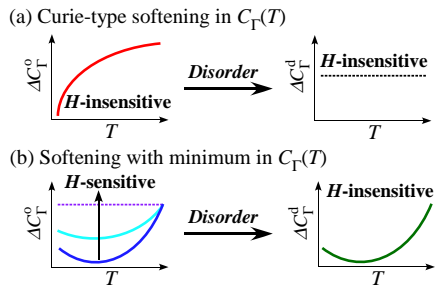}
\caption{\label{fig:Fig3} (Color online). Schematic of the two-types of the elastic anomalies in the cubic PM phase of MgV$_2$O$_4$. (a) Disorder-sensitive Curie-type softening in $C_{\Gamma}(T)$ and (b) $H$- and disorder-sensitive softening with minimum in $C_{\Gamma}(T)$.}
\end{center}
\end{figure}

We remark here that the component of the softening with minimum in $C_{44}^o(T)$ and $C_{11}^o(T)$ respectively seen in Figs. 2(c) and 2(a) is sensitive to $H$, whereas that in $C_{\Gamma}^d(T)$ seen in Figs. 2(d)-2(f) is insensitive to $H$, as illustrated in Fig. 3(b). This difference in the $H$-sensitivity should also arise due to disorder, where the introduction of disorder causes a change in the nature of the magnetic excitations. The detailed mechanism for this disorder effect remains to be elucidated. For instance, a disorder-induced change from the orbital-spin-coupled excitations to the orbital-only excitations might occur in MgV$_2$O$_4$.

As reported in Ref. [\cite{Nii}], $C_{\Gamma}(T)$ of MnV$_2$O$_4$ exhibits only the Curie-type softening in all the elastic modes, indicating that the component of the softening with minimum in $C_{\Gamma}(T)$ is absent or negligibly small. From the discussion on the orbital order in MgV$_2$O$_4$, it is probable that the effect of the spin-orbit coupling in MgV$_2$O$_4$ is stronger than that in MnV$_2$O$_4$ \cite{Wheeler}. Thus implying that the spin-orbit coupling plays a crucial role for the occurrence of the softening with minimum in $C_{\Gamma}(T)$ of MgV$_2$O$_4$.

In summary, ultrasound velocity measurements of MgV$_2$O$_4$ reveal the coexisting two-types of elastic anomalies in the cubic PM phase which are resolved by sensitivity to $H$ and disorder, as illustrated in Figs. 3(a) and 3(b). These elastic anomalies can be attributed to the coexistence of orbital fluctuations and orbital-spin-coupled excitations. The present results suggest that the geometrical frustration, and the interplay of spin, orbital, and lattice degrees of freedom, evoke complex dynamical phenomena in the cubic PM phase of $A$V$_2$O$_4$ which deserve further experimental and theoretical studies.

We thank K. Tomiyasu and K. Kamazawa for stimulating discussions. This work was partly supported by Grant-in-Aid for Scientific Research (C) (25400348) from MEXT of Japan, and by Nihon University College of Science and Technology Grants-in-Aid for Fundamental Science Research.

\end{document}